\documentclass{ws-mpla}
\usepackage[super]{cite}
\usepackage{graphicx}
\newcommand{\be}{\begin{equation}}
\newcommand{\ee}{\end{equation}}
\newcommand{\OL}{\Omega_{\Lambda}}

\begin{document}

\markboth{Spyros Basilakos}
{Dynamics in $\Lambda(H)$ cosmologies}

\catchline{}{}{}{}{}

\title{Cosmic expansion and structure formation in running vacuum cosmologies}


\author{\footnotesize Spyros Basilakos\footnote{
email: svasil@academyofathens.gr}}

\address{Research Center for Astronomy and Applied Mathematics, Academy of Athens \\
Soranou Efesiou 4, 11527, Athens, Greece
\\
svasil@academyofathens.gr}

\maketitle


\begin{abstract}
We investigate the dynamics of the FLRW flat cosmological models 
in which the vacuum energy varies with redshift.
A particularly well motivated model
of this type is the so-called quantum field vacuum, in which
both kind of terms $H^{2}$ and constant appear  
in the effective dark energy density affecting the 
evolution of the main cosmological functions at the background and
perturbation levels.
Specifically, it turns out 
that the functional form of the quantum vacuum endows the vacuum energy of a 
mild dynamical evolution which could be observed nowadays 
and appears as dynamical dark energy. Interestingly, 
the low-energy behavior is very close to the 
usual $\Lambda$CDM model, but it is by no means identical. 
Finally, within the framework of the 
quantum field vacuum 
we generalize the large scale structure properties, namely 
growth of matter perturbations, cluster number counts and 
spherical collapse model.

\keywords{Cosmology; Dark Energy; Large Scale Structure.}
\end{abstract}

\ccode{PACS Nos.: 98.80.-k, 95.35.+d, 95.36.+x}

\section{Introduction}
The statistical analysis of various cosmological data
(SNIa, Cosmic Microwave Background-CMB, Baryonic Acoustic
Oscillations-BAOs, Hubble parameter measurements etc) strongly
suggests that we live in a spatially flat universe
that consists of $\sim 4\%$ baryonic matter, $\sim 26\%$
dark matter and $\sim 70\%$ some sort of dark energy (hereafter DE) which is
necessary to explain the accelerated expansion of the
universe (see Refs.[\refcite{Hicken2009,PlanckXVI2013}] and references
therein).
Although there is a common agreement regarding the
ingredients of the universe, there are different views
concerning the possible physical mechanism which is responsible
for the cosmic acceleration. 

The simplest dark energy candidate corresponds to a cosmological
constant. In the standard concordance $\Lambda$CDM model, the
overall cosmic fluid contains baryons, cold dark matter plus a
vacuum energy (cosmological constant), 
that appears to fit accurately the current observational data
and thus provides an excellent scenario to describe the observed
universe. However, the concordance model suffers from, among other, 
two fundamental problems: 
(a) {\it The fine tuning problem} i.e., the fact 
that the observed value of the
vacuum energy density ($\rho_{\Lambda}=\Lambda c^{2}/8\pi G\simeq
10^{-47}\,GeV^4$) is many orders of magnitude below the value found
using quantum field theory (QFT)\cite{Weinberg1989}, and (b) {\it the
coincidence problem}\cite{Steinhardt1997}, 
i.e., the fact
that the matter energy density and the vacuum energy density are of
the same order just prior to the present epoch, despite
the fact that the former is a rapidly decreasing function
of time while the latter is stationary. 
Such problems have inspired many authors to propose
alternative candidates to the concordance $\Lambda$CDM, such 
as general dark energy models and modifications of the theory of
gravity (for review see Ref.[\refcite{Amendola2010}]).

An alternative path that one can follow
in order to alleviate the above problems is to consider 
a time varying vacuum, $\Lambda(t)$. Within this framework 
we do not need to include new fields in nature nor
to modify the theory of General Relativity.
In this cosmological scenario the dark energy equation
of state parameter $w\equiv P_{DE}/\rho_{DE}$,
is strictly equal to -1, but the vacuum energy density 
is not a constant but rather it is a function of the cosmic 
time. 
Note that there is an extensive (old\cite{old,Arc94}  
and new\cite{Ber,Shap00,RGTypeIa,BPS09,Grande,Andria}) 
literature in which the time-evolving vacuum has 
been phenomenologically modeled as a function of time in various
possible ways, in particular, as a function of the Hubble
parameter.

In this article we reconsider the so called quantum field vacuum (see next
section). We would like to stress that this vacuum 
model was previously investigated by our team in many aspects\cite{}.
The layout of the article is the following. In section 2, we
present the main cosmological ingredients of the quantum field vacuum model.
In section 3 we provide the power spectrum 
and the linear growth of matter perturbations.
In sections 4 and 5 we discuss the cluster number counts and the 
spherical collapse model. 
Finally, we draw our conclusions in section 6.

\section{Background expansion - Observational constraints}
The nature of the current vacuum model\cite{Shap00,RGTypeIa,BPS09,Grande} is essentially connected with the renormalization group (RG) 
in quantum field theory (QFT). In this context, the evolution of the 
vacuum is written as 
\be 
\label{RGlaw2}
\Lambda(H)=\Lambda_0+ 3\nu\,(H^{2}-H_0^2)\,.
\ee
where $\Lambda_0\equiv\Lambda(H_0)=3\Omega_{\Lambda}H^{2}_{0}$ and 
$\nu$ is interpreted in the RG
framework as a ``$\beta$-function'' of QFT in curved spacetime,
which determines the running of the cosmological constant. 
Regarding the dynamical role of $\nu$ 
it has been found that  
since $|\nu|\ll 1$ (for a recent 
discussion see Ref.~[\refcite{Andria}] and references therein)
at low redshifts the model becomes almost indistinguishable
from the cosmic concordance model. In other words 
the $\nu$-parameter endows the vacuum energy of a 
mild dynamical evolution which could be observed nowadays 
and appears as dynamical dark energy. The low-energy behavior 
is thus very close to the concordance model, but it 
is by no means identical.

Naturally, the next step here is to 
derive the Friedmann equations assuming 
of course a running vacuum.
This procedure
is perfectly allowed by the Cosmological Principle
embedded in the FLRW metric. 
In this context the Friedmann 
equations are written
\be
 8\pi G\rho_{\rm tot}\equiv 8\pi G \rho_{m}+\Lambda =
3\left(\frac{\dot a}{a}\right)=3H^{2}\;,
\label{friedr} 
\ee
\be 8\pi G p_{\rm tot} \equiv   8\pi G p_{m}-\Lambda =-2{\dot H}-3H^2
\label{friedr2} 
\ee 
where the overdot denotes derivative with respect to
cosmic time $t$. 
Notice that the Bianchi 
identities 
that insure the covariance of the theory, pose an energy exchange between 
vacuum and matter for $G=const.$ 
\begin{equation}
\dot{\rho}_{m}+3(1+\omega_{m})H\rho_{m}=-\dot{\rho_{\Lambda}}\,. \label{frie33}
\end{equation}
Combining equations (\ref{friedr}), (\ref{friedr2}) 
and (\ref{frie33}), we provide the basic 
differential equation that governs the dynamics of the Universe
\begin{equation}
\dot{H}+\frac{3}{2}(1+\omega_{m}) H^{2}=
\frac{(1+\omega_{m})}{2}\Lambda(H) \;.
\label{frie34}
\end{equation}
Since we are in the matter era we set $\omega_{m}\equiv 0$.

Inserting eq.(\ref{RGlaw2}) into eq.(\ref{frie34}) we can integrate 
the latter in order to obtain 
the Hubble parameter as a function of time:
\begin{equation}
\label{frie455} H(t)=H_{0}\,\sqrt{\frac{\OL-\nu}{1-\nu}} \;
\coth\left[\frac32\,H_{0}\sqrt{(\OL-\nu)(1-\nu)}\;t\right]\,.
\end{equation}
where $\Omega_{\Lambda}=1-\Omega_{m}$ and 
$H_{0}$ is the Hubble constant\footnote{For the comoving distance
and for the dark matter halo mass (see section 4) we use the traditional
parametrization $H_{0}=100h$km/s/Mpc. Of course, when we treat the
power spectrum shape parameter $\Gamma$ we utilize
$h\equiv h_{\rm Planck}=0.673$ [\refcite{PlanckXVI2013}].}.
Using $H={\dot a}/a$ the cosmic time, $t(a)$, is given by
\begin{equation}
t(a)=\frac{2}{3\,{\tilde{\Omega}_{\Lambda}}^{1/2}\,(1-\nu)\,H_{0}} 
{\rm sinh^{-1}} \left(\sqrt{ \frac{\tilde{\Omega}_{\Lambda}}
{\tilde{\Omega}_{m}}} \;a^{3(1-\nu)/2} \right) \label{frie456t}
\end{equation}
where we have introduced
\begin{equation}
\label{otran1}
\tilde{\Omega}_{m}=\frac{\Omega_{m}}{1-\nu}\,,\ \ \
\tilde{\Omega}_{\Lambda}=\frac{\Omega_{\Lambda}-\nu}{1-\nu}\,.
\end{equation}
Inverting eq.(\ref{frie456t}) we easily determine the scale factor $a=a(t)$.
Substituting the cosmic time into (\ref{frie455}),
one can provide the normalized Hubble parameter
\begin{equation}
\label{anorm11}
E^{2}(a)=\frac{H^{2}(a)}{H^{2}_{0}}=
\tilde{\Omega}_{\Lambda}+\tilde{\Omega}_{m}a^{-3(1-\nu)}\,
\end{equation}
where the cosmological parameters obey the 
standard cosmic sum rule, namely
$\tilde{\Omega}_{m}+\tilde{\Omega}_{\Lambda}=1=\Omega_{m}+\Omega_{\Lambda}$.

Concerning the evolution of matter density the situation is as follows.
Combining eq.(\ref{RGlaw2}), (\ref{friedr}) and  
(\ref{frie33}) we arrive at 
$\dot{\rho}_m+3H\rho_m=3\nu H\rho_{m}$.
Integrating the latter (using $\dot{\rho}_m=aH d\rho_m/da$)
we find
\begin{equation}\label{mRG}
\rho_m(a) =\rho_{m0}\,a^{-3(1-\nu)}\,,
\end{equation}
where $\rho_{m0}$ is the matter density at the present time ($a=1$),
and therefore $\Omega_m=\rho_{m0}/\rho_{c0}$, where
$\rho_{c0}=3H_0^2/8\pi G$ is the current critical density.
In fact, defining $\Omega_m(a)\equiv{\rho_m(a)}/{\rho_c(a)}$ it is
easy to see, with the help of (\ref{mRG}) and the definition of
$E(a)$, that
\begin{equation}\label{effeom}
\Omega_m(a)=\frac{\Omega_{m}a^{-3(1-\nu)}}{E^2(a)}\,.
\end{equation}
Lastly, upon inserting (\ref{mRG}) in (\ref{frie33}) and integrating
once more in the scale factor variable, we arrive at the explicit
expression for the evolution of the vacuum energy density:
\begin{equation}
\label{CRG}
\Lambda(a)=\Lambda_0+8\pi G \;\frac{\nu\,\rho_{m0}}{1-\nu}\,
\left[a^{-3(1-\nu)}-1\right]\,.
\end{equation}
Obviously, for $\nu=0$ the current time varying vacuum model 
reduces to the concordance 
$\Lambda$ cosmology as it should.	

Using the above normalized Hubble 
parameter [see eq.(\ref{anorm11})], a joint statistical analysis, involving 
the latest cosmological data [SNIa\cite{Suzuki}, $A(z)$ of BAOs\cite{Blake11}
and {\it Planck} CMB shift parameter \cite{PlanckXVI2013,Shaf13}] 
is implemented. 
Since the explored cosmological models ($\Lambda_{RG}$ and $\Lambda$CDM) 
contain different number of free
parameters, as a further statistical test we use the
({\em corrected}) Akaike Information Criterion (AIC)
relevant to our case ($N_{\rm tot}/k>40$)\cite{Akaike1974}, which 
is given, in the case of Gaussian errors, as follows: 
${\rm AIC}=\chi^2_{\rm t, min}+2k$, 
where $\chi^{2}_{\rm t}=\chi^{2}_{\rm SNIa}+\chi^{2}_{\rm BAO}+\chi^{2}_{\rm CMB}$ 
is the overall $\chi^2$-function and 
$k$ is the number of free parameters. From the statistical viewpoint,
a smaller value of AIC points to a better model-data fit. Within this 
framework, we need to make clear that small
differences in AIC are not necessarily significant and therefore, it is
important to derive the model pair difference, namely
$\Delta$AIC$ = {\rm AIC}_{y} - {\rm AIC}_{x}$. The larger the
value of $|\Delta{\rm AIC}|$, the higher the evidence against the
model with larger value of ${\rm AIC}$, with a
difference $|\Delta$AIC$| \ge 2$ indicating a positive
such evidence and $|\Delta$AIC$| \ge 6$
indicating a strong such evidence, while a value $\le 2$ indicates
consistency among the two comparison models.
For the concordance $\Lambda$CDM cosmology, if we 
impose $\nu=0$ and minimizing with respect to $\Omega_{m}$
we find $\Omega_{m}=0.291\pm 0.011$ with
$\chi_{\rm t, min}^{2}(\Omega_{m})/dof \simeq 567.5/586$
(AIC$_{\Lambda}$$\simeq 569.5$). For comparison, we provide the results 
of {\it Planck}: $\Omega_{m}\,h^2=0.1426\pm 0.0025$, with  
$h_{\rm Planck}=0.673\pm 0.012$ which implies 
$\Omega_{m}=0.315\pm 0.016$. Notice that throughout the paper 
we set $\Omega_{m,\Lambda}\equiv \Omega_{m}=0.291$. 
In the case of the $\Lambda_{RG}$ model we find that the overall
likelihood function peaks at $\Omega_{m}=0.282\pm 0.012$ and
$\nu=0.0048\pm 0.0032$ with 
$\chi_{\rm t, min}^{2}(\Omega_{m},\nu) \simeq 563.8$ (AIC$_{\rm RG}$$\simeq 567.8$)
for $585$ degrees of 
freedom\cite{Andria}.
It turns out that the
current vacuum model appears to
fit slightly better than the $\Lambda$CDM the
observational data.
Still, the $|\Delta {\rm AIC}|$=$|{\rm AIC}_{\rm \Lambda}-{\rm AIC}_{RG}|$
values (ie., $\le 2$) indicate that
the cosmological data are simultaneously consistent with the 
cosmological models.
Note that for the rest of the
paper we will restrict our present analysis to the choice
of $(\Omega_{m},\nu)=(0.282,0.0048)$.

\begin{table}[h]
\tbl{Statistical results from fitting SNIa+BAO+CMB$_{shift}$ data.
The $1^st$ column indicates the model. The $2^{nd}$ and $3^{nd}$ columns 
provide the best fit 
parameters $(\Omega_{m},\nu)$. The $4^{nd}$ and $5^{nd}$ columns 
list the statistical significance of the fit. Finally, the $6^{th}$ and 
$7^{th}$ columns shows the estimated values of $(\sigma_{8},\delta_{c})$. Notice 
that in section 4 we provide a relevant discussion regarding $\delta_{c}$.}
{\begin{tabular}{@{}ccccccc@{}} \toprule
Model & $\Omega_{m}$ & $\nu$ &
$\chi^{2}_{\rm t,min}$ & {\rm AIC}&$\sigma_{8}$&$\delta_{c}$ \\
\colrule
$\Lambda$CDM& $0.291\pm 0.011$& 0 & 567.5 & 569.5 & 0.829&1.675 \\
$\Lambda_{RG}$& $0.282\pm 0.012$& $0.0048\pm0.0032$ & 563.8 & 567.8 & 0.758& 1.644 \\
\end{tabular}\label{ta1} }
\end{table}

\section{CDM Power Spectrum - growth factor}
The CDM power spectrum is given by $P(k)=P_{0} k^{n}T^{2}(k)$,
where $T(k)$ is the CDM transfer function
and $n\simeq 0.9603$ following the recent analysis of the Planck
data \cite{PlanckXVI2013}. Concerning the form of $T(k)$,
we use that of Eisenstein \& Hu \cite{Eisenstein1998}
$T(k)=\frac{L_{0}}{L_{0}+C_{0}q^{2}}$, 
where $q\equiv \frac{k}{\Gamma}$,
$L_{0}={\rm ln}(2e+1.8q)$, $e=2.718$ and $C_{0}=14.2+\frac{731}{1+62.5q}$.
Here $\Gamma$ is the shape parameter
\cite{Sugiyama1995} which is written as:
$\Gamma=\Omega_{m}h_{\rm Planck}\;{\rm
  exp}(-\Omega_{b}-\sqrt{2h_{\rm Planck}}\;\Omega_{b}/\Omega_{m})$.
The value of $\Gamma$, which is kept constant throughout the model
fitting procedure, is estimated using
the {\it Planck} results
$\Omega_{b}=0.0222h^{-2}_{\rm Planck}$ with $h_{\rm Planck}=0.672$. 

Another important quantity here is the rms fluctuations of the linear density
field on mass scale $M_{h}$:
\be \label{s888} \sigma^2(M,z)=\sigma^2_8(z)
\frac{\int_{0}^{\infty} k^{n_s+2} T^{2}(\Omega_{m}, k) W^2(kR)
dk} {\int_{0}^{\infty} k^{n_s+2} T^{2}(\Omega_{m}, k)
W^2(kR_{8}) dk},\;\;\; \sigma_8(z)=\sigma_8\frac{D(z)}{D(0)} \;. 
\ee 
where $\sigma_{8}\equiv \sigma(R_{8},0)$ is the rms mass fluctuation
on $R_{8}=8 h^{-1}$ Mpc scales. 
At this point we would like to remind the reader that 
$D(z)$ is the growth factor 
(see below), $W(kR)=3({\rm sin}kR-kR{\rm cos}kR)/(kR)^{3}$ and
$R=(3M_{h}/ 4\pi \rho_{m0})^{1/3}$ with
$\rho_{m0}=\Omega_{m}\rho_{c0}$ denotes the mean matter density of the
universe at the present time
($\rho_{m0}=2.78 \times 10^{11}\Omega_{m}h^{2}M_{\odot}$Mpc$^{-3}$).
In this work we use the {\it Planck} prior, namely 
$\sigma_{8,\Lambda}=0.829$. Of course, in order to use $\sigma_{8}$
properly along the current vacuum model we need to rescale the value of   
$\sigma_{8}$ by 
\be \label{s88general}
\sigma_{\rm 8}=\sigma_{8, \Lambda} \frac{D(0)}{D_{\Lambda}(0)}
\left[\frac{P_{0}\int_{0}^{\infty} k^{n_s+2}
T^{2}(\Omega_{m},k) W^2(kR_{8}) dk}
{P_{0,\Lambda}\int_{0}^{\infty} k^{n_s+2} T^{2}(\Omega_{m,
\Lambda},k) W^2(kR_{8}) dk} \right]^{1/2}\,,
\ee
where $P_{0}/P_{0,\Lambda}=(\Omega_{m,\Lambda}/\Omega_{m})^{2}$. 
Based on the aforementioned observational constraints and eq.(\ref{s88general})
we find $\sigma_{8}=0.758$ for the $\Lambda_{RG}$ model. 
In Table~1, one may see a more compact presentation of the cosmological 
parameters including the $\delta_{c}$ which is the
linearly extrapolated density
threshold above which structures collapse (see section 4).

Now we focus on 
the basic equation which describes the evolution of matter
fluctuations within the context of the previously described vacuum 
model. 
At sub-horizon scales the corresponding time
evolution equation (in the linear regime) for the matter density contrast
$D\equiv\delta\rho_m/\rho_m$, in a pressureless fluid, is given
by (see Refs. [\refcite{Arc94,Borg08}]):  
\begin{equation}
\label{eq:11} 
\ddot{D}+(2H+Q)\dot{D}-\left(4\pi G \rho_{m}
-2HQ-\dot{Q} \right)D=0,
\end{equation}
where in our case $Q(t)=-\dot{\Lambda}/\rho_{m}$.
It becomes clear, that the
interacting vacuum energy affects the growth factor via the function
$Q(t)$. In the case of non interacting 
DE models, [$Q(t)=0$], the
above equation (\ref{eq:11}) reduces to the usual time evolution
equation for the mass density contrast
\cite{Peeb93,Linjen03}. 
For the concordance $\Lambda$ cosmology it is easy to prove that 
the solution of eq.(\ref{eq:11}) 
is \cite{Peeb93} 
\be\label{eq24}
D_{\Lambda}(a)=\frac{5\Omega_{\rm m}
  E(a)}{2}\int^{a}_{0}
\frac{dx}{x^{3}E^{3}(x)} \;\;. 
\ee

\begin{figure}[ph]
\centerline{\includegraphics[width=5.0in]{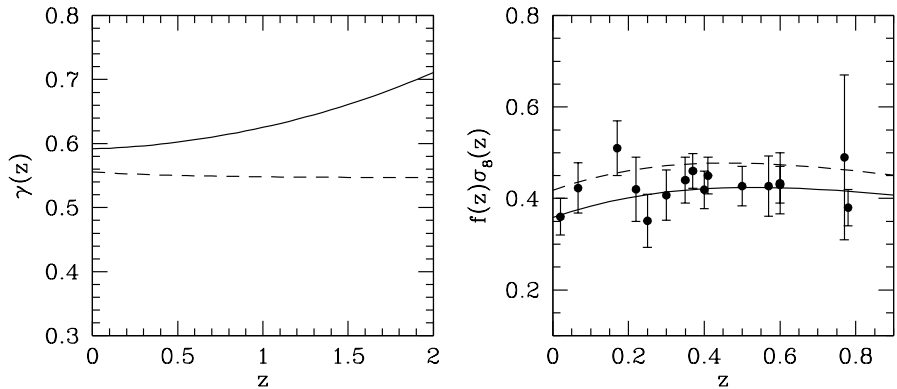}}
\vspace*{8pt}
\caption{{\em Right Panel:} Comparison of the observed and
theoretical evolution of the growth
rate $f(z)\sigma_{8}(z)$.
The dashed line corresponds to the
$\Lambda$CDM ($\sigma_{8,\Lambda}=0.829$) and the solid one is for 
$\Lambda_{RG}$ ($\sigma_{8}=0.758$). 
{\em Left Panel:} The evolution of the growth index of matter 
perturbations.\protect\label{fig1}}
\end{figure}
For the purpose of our study it is important to solve 
eq.(\ref{eq:11}) in order to investigate the matter
fluctuation field of the vacuum model (\ref{RGlaw2}) in the linear
regime. Here we present the main steps (for more details 
see Ref.~[\refcite{BPS09}]).
First of all we change variables from $t$ to a new one
according to the transformation
\be\label{tran11} y={\rm
coth}\left[\frac{3}{2}\,H_{0}\sqrt{(\Omega_{\Lambda}-\nu)\,(1-\nu)}\;t\right]
\;\;. \ee
Also from equations (\ref{frie455}), (\ref{frie456t}) and (\ref{anorm11})
we get the following useful relations:
\begin{equation}
\label{pps11}
y=\sqrt{\frac{1-\nu}{\Omega_{\Lambda}-\nu}}\,E(a)
\;,\;\;\;\;\;\;
y^{2}-1=\frac{\Omega_{m}}{\Omega_{\Lambda}-\nu}\;a^{-3(1-\nu)}\,.
\end{equation}
Utilizing (\ref{tran11}) and
(\ref{pps11}) we find, after some non-trivial algebra, that equation
(\ref{eq:11}) becomes
\be \label{eq:222} 3\beta^2 (y^{2}-1)^{2}D^{''}+k(\beta)
y(y^{2}-1)D^{'}- 2[g(\beta)y^{2}-\psi(\beta)]D=0\,, 
\ee
a solution of which is 
\begin{equation}
\label{DRG}
D(y)={\cal C}(y^{2}-1)^{\frac{4-9\beta}{6\beta}} y F\left(\frac{1}{3\beta}+
\frac{1}{2},\frac{3}{2},
\frac{1}{3\beta}+\frac{3}{2},-\frac{1}{y^{2}-1} \right)
\end{equation}
where $\beta\equiv 1-\nu$,  
primes denote derivatives with respect to $y$, the quantity $F$ is the 
hypergeometric function,
$k(\beta)=2\beta(6\beta-5)$, 
$g(\beta)=(2-\beta)(3\beta-2)$ and 
$\psi(\beta)=\beta(4-3\beta)$.
Now inserting
eq.(\ref{pps11}) into eq.(\ref{DRG}) and using eq.(\ref{otran1}), we
finally obtain the growth factor $D(a)$ as a function of the scale
factor:
\begin{equation}
\label{DRG1}
D(a)=C_{1} a^{\frac{9\beta-4}{2}}E(a)
F\left(\frac{1}{3\beta}+\frac{1}{2},\frac{3}{2},
\frac{1}{3\beta}+\frac{3}{2},-\frac{\tilde{\Omega}_{\Lambda}}{\tilde{\Omega}_{m}}\,\,a^{3\beta}
\right)
\end{equation}
where $C_{1} \propto {\cal C}$ is an integration constant to be adjusted 
by an initial condition.\footnote{We normalize the growth factor 
at large redshifts ($z\gg 1$), namely $D(z)\simeq 1/(1+z)$.} 

Furthermore, a crucial role in structure formation studies plays the 
growth rate of clustering which is defined as
$f(a)=\frac{d{\rm ln}D}{d{\rm ln}a} \simeq \Omega_{m}^{\gamma}(a)$
where $\gamma$ is the growth index.
It has been shown that for those dark energy 
models which have a slow varying equation of state parameter, the 
growth index $\gamma$ is 
well approximated by $\gamma \simeq \frac{3(w-1)}{6w-5}$
(see Refs.~[\refcite{Silv1994,Wang1998,Nesseris2008}]), or 
$\gamma\simeq 6/11$ in the case where $w=-1$ ($\Lambda$CDM model).
Now since we know the analytical form of the growth factor $D(a)$ one may 
directly compute the evolution of $f(a)$.

In order to test the performance of the 
$\Lambda_{RG}$ vacuum model at the perturbation level, 
we utilize the recent growth rate data
for which their combination parameter of the growth rate of structure,
$f(z)$, and the redshift-dependent rms fluctuations of the linear
density field, $\sigma_8(z)$,
is available as a function of redshift, $f(z)\sigma_{8}(z)$.
The total sample contains $N=16$ entries
(as collected by Ref.~[\refcite{BasilakosNes2013}] - see their Table 1 
and references therein).
In the right panel of figure 1 we present the growth data (solid points) 
together with the predicted $f(z)\sigma_{8}(z)$
for the running vacuum model (solid line) and $\Lambda$CDM (dashed line). 
We find that the concordance $\Lambda$ cosmology (with $\Omega_{m}=0.291$)
reproduce the growth data with $\chi_{\rm min}^{2}/dof \simeq 20.02/15$ implying
that the $\Lambda$CDM model can not simultaneously accommodate the 
{\em Planck} priors and the growth data (see also Ref.~[\refcite{Mac13}]).
On the other hand for the $\Lambda_{RG}$ model we have 
$\chi_{\rm min}^{2}/dof \simeq 8.42/15$. 
Performing the AIC information criterion for small sample
size, namely ${\rm AIC} =\chi^{2}_{\rm min}+2k+\frac{2k(k-1)}{N-k-1}$
we find $\Delta$AIC$= {\rm AIC}_{\Lambda}-{\rm AIC}_{RG} \simeq 11.6$
which means that the growth data favor the current running vacuum 
scenario.

Let us finish this section with a relevant discussion regarding 
the growth index $\gamma$. As we have already 
described, we can express the linear growth rate of clustering
in terms of the growth index. Specifically, from $\frac{d{\rm ln}D}{d{\rm ln}a} \simeq \Omega_{m}^{\gamma}(a)$
we easily obtain:
$\gamma(z)\simeq \frac{\ln\left[-(1+z) \frac{d\ln D}{dz}\right]}
{\ln\Omega_{m}(z)}$,
where $z=a^{-1}(z)-1$. 
Based on the aforesaid observational constraints 
we find that the growth index at the present time becomes 
$\gamma_{\Lambda_{RG}} \simeq 0.58$ which is somewhat higher with respect to the 
$\Lambda$CDM value, namely $\gamma \simeq 6/11$.
In the left panel of figure 1 we show
the growth index evolution for 
the $\Lambda_{RG}$ and $\Lambda$CDM models, respectively.
From the comparison we observe that the growth index of the 
$\Lambda_{RG}$ vacuum model deviates from the concordance model.
In particular, there 
is a visible deviation from above in all 
the redshift range. This deviation becomes at the level of $10\%$ 
for $z \le 1$. 

\begin{figure}[ph]
\centerline{\includegraphics[width=5.0in]{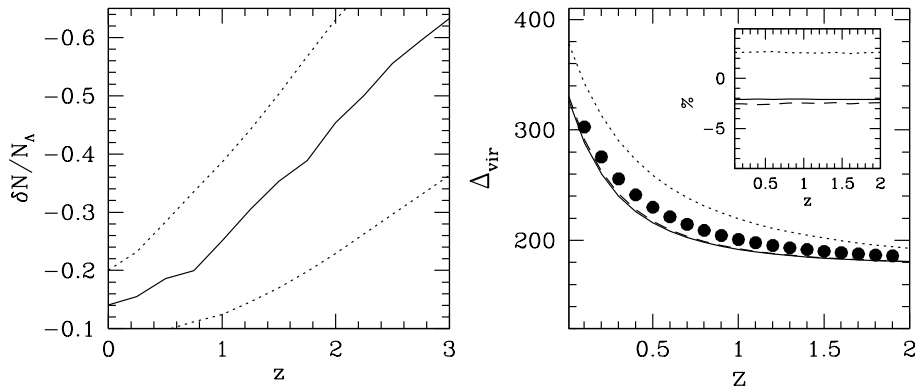}}
\vspace*{8pt}
\caption{{\it Left Panel:} Fractional 
difference $\delta{\cal N}/N_{\Lambda}$ (solid line) 
in the cluster
number counts between the $\Lambda_{RG}$ model and the reference
$\Lambda$CDM model. 
The dotted curves correspond to 2$\sigma$ Poisson
uncertainties. Notice that we use the priors of Table 1.
{\it Right Panel:} In the inner panel we provide 
the relative deviation $(1-\lambda/\lambda_{\Lambda})\%$
of the collapse factor for various vacuum models with respect to the
$\Lambda$ solution. In the outer panel we observe the evolution 
of the virial density.
The lines corresponds to the following vacuum models: (i)
$\Lambda_{RGH}$ (solid), (ii)
$\Lambda_{RGC1}$ (dashed line, $\nu_{s}=0.002$) and (iii)
$\Lambda_{RGC2}$ (dotted line, $\nu_{s}=-0.002$).
Notice that the solid points represent the concordance 
$\Lambda$CDM cosmology.
\protect\label{fig2}}
\end{figure}

\section{Comparison with Cluster Halo Abundances}\label{sec:cluster}
Another choice in order to discriminate the models, that 
has been extensively used so far in the literature, is to
estimate the theoretical predictions of the models for 
the cluster-size halo redshift distributions and to confront them 
with the data. 
The halo abundances predicted by a 
large variety of time varying vacuum models have been compared 
with those corresponding
to the $\Lambda$CDM model~\cite{BPS09,Grande,Andria,Alc14}. 
We utilize the Press and Schechter~\cite{press} formalism,
which determines the fraction of matter that has formed bounded
structures as a function of redshift. Mathematical details of such a
treatment can be found also in Ref.~[\refcite{Andria}]; here we 
only present the basic ideas.
The number density of halos, $n(M,z)$,
with masses within the range $(M,\:M+\delta M)$ is given by:
\begin{equation}
\label{MF}
n(M,z) dM = \frac{\bar{\rho}}{M} \frac{d{\rm \ln}\sigma^{-1}}{dM} f(\sigma) dM \;,  
\end{equation}
where $\bar\rho=\rho_{m0}$ is the mean background mass density (see section 3). 
In the original Press-Schechter (PSc) formalism,
$f(\sigma)=f_{\rm PSc}(\sigma)=\sqrt{2/\pi} (\delta_c/\sigma)
\exp(-\delta_c^2/2\sigma^2)$, 
$\delta_{c}$ is the
linearly extrapolated density
threshold above which structures collapse~\cite{eke}, while
$\sigma^2(M,z)$ is the mass variance of
the smoothed linear density field, extrapolated to redshift $z$ at which
the halos are identified [see eq.(\ref{s888})]. 
The mass variance depends on the power-spectrum of
density perturbations in Fourier space, $P(k)$, for which we use here the
CDM form according to Ref.~[\refcite{Eisenstein1998}], and the 
values of the
baryon density parameter, the spectral slope and Hubble constant
according to the recent {\it Planck} results \cite{PlanckXVI2013}.
Although the original Press-Schechter mass-function, $f_{\rm PSc}$,
was shown to provide a good first approximation to that
provided by numerical simulations, it was later found
to over-predict/under-predict the number of low/high mass halos at the
present epoch \cite{Jenk01,LM07}.
More recently, a large number of works have provided better fitting
functions of $f(\sigma)$, some of them  based on a phenomenological
approach. In the present
study, we adopt the one proposed by Reed \emph{et al}. [\refcite{Reed}].

In order to compare the mass function predictions of the
different cosmological models, it is important to use
for each model the corresponding value of $\sigma_8$ and $\delta_c$ 
(see Table 1). 
It is well known that for the usual $\Lambda$ cosmology
we have $\delta_{c} \simeq 1.675$ (see Ref.[\refcite{Wein03}]). 
For the $\Lambda_{RG}$ model it has been found 
that $\delta_{c} \simeq 1.644$ (see Ref.~[\refcite{Andria}]). 
Given the halo mass function from eq.(\ref{MF}) we can 
now derive an observable
quantity which is the redshift distribution of clusters, ${\cal
N}(z)$, within some determined mass range, say $M_1\le
M/h^{-1}M_{\odot}\le M_2=10^{16}$. This can be estimated by integrating over mass
the expected differential halo mass function, $n(M,z)$:
\be {\cal
N}(z)=\frac{dV}{dz}\;\int_{M_{1}}^{M_{2}} n(M,z)dM,\;\;\;\frac{dV}{dz}=\Omega_{s}r^{2}(z)\frac{dr}{dz},\;\;\;\; r(z)=c \int_{0}^{z} \frac{du}{H(u)}\;,
\ee
where $dV/dz$ is the comoving volume element
and $\Omega_{s}$ is the solid angle.

In left panel of figure 2, we present the fractional difference 
between the $\Lambda_{RG}$ model and $\Lambda$CDM
namely, $\delta {\cal N}/{\cal N}_{\Lambda}$, 
with masses in the range  
$10^{13.4}\,h^{-1}\lesssim M/M_{\odot}\lesssim 10^{16}\,h^{-1}$.
Recall that we denote the deviations of the number 
counts of a given vacuum model 
with respect to the $\Lambda$CDM as 
$\delta\mathcal{N}=\mathcal{N}-\mathcal{N}_{\Lambda}$.
The dotted lines shown in the left panel of figure 2 
correspond to 2$\sigma$ Poisson
uncertainties, which however do not include cosmic variance
and possible observational systematic uncertainties, that would
further increase the relevant variance.
Specifically, the prediction for  
$\delta {\cal N}/{\cal N}$ is negative for all points, even 
for those in the $\pm 1\sigma$ band. 
Thus, the running vacuum model tends 
to predict a smaller number of counts as compared 
to the $\Lambda$CDM. 
The fractional decrease can be as significant as $30-60\%$.
We also find 
that the redshift variation
of the differences between the $\Lambda_{RG}$ cosmology and 
$\Lambda$CDM model is mainly affected by variations in
the values of $\sigma_{8}$ and $\delta_{c}$ 
(for more discusion see Ref.~[\refcite{Andria}]). 
Therefore, we have verified that there are observational signatures
that can be used to differentiate the $\Lambda_{RG}$ model from the
$\Lambda$CDM and possibly from a large class of DE models.

\section{The spherical collapse model}
In this section we 
generalize the spherical collapse model 
within the variable $\Lambda(H)$ cosmological
model, in order to understand non-linear structure formation in
such scenarios and investigate the differences with the corresponding
expectations of the concordance $\Lambda$CDM cosmology.
Practically, one may start from the Raychaudhuri equation
which is valid either for 
the entire universe or for homogeneous spherical perturbations 
[by replacing the scale factor with radius $R(t)$] 
\begin{equation}
\frac{\ddot{R}}{R}=-\frac{4\pi G}{3}\left(\rho_{ms}-2\rho_{\Lambda
s}\right)\,,\label{rayc}
\end{equation}
where $\rho_{ms}$ and $\rho_{\Lambda s}$ refer to the
corresponding values of the matter and vacuum energy densities in
the spherical patch susceptible of ulterior collapse.

Now let us first define some basic quantities of the problem. In particular, 
we call $\alpha_{t}$ the scale factor of the
universe where the overdensity reaches 
at its maximum expansion (i.e. when $\dot R=0$) and 
$\alpha_{c}$ the scale factor in which the sphere virializes, 
implying that a cosmic structure has formed.
Also, $R_{t}$ and $R_{c}$ denote the corresponding radii of the
spherical overdensity, the former being the turnaround (or
``top hat'') value at the point of maximum size, and the latter
refers to the eventual situation when the sphere has already
collapsed and virialized. We would like to remind the reader that 
due to the coupling between the running vacuum and matter
one would expect that
the matter density in the spherical overdensity should obey the same
power law as the background matter $\rho_{m}(a)\propto
a^{-3(1-\nu)}$ (see eq.\ref{mRG}). Thus, $\rho_{ms}\propto
R^{-3(1-\nu)}$ denotes the matter density in the spherical patch.
Regarding, the vacuum energy density in the spherical
region, $\rho_{\Lambda s}$ we focus on 
two situations: 
(i) the vacuum
energy remains homogeneous and only the corresponding matter
virializes (this scenario holds for the $\Lambda$ cosmology); and 
(ii) the
case with clustered vacuum
energy, assuming that the whole system virializes (both
matter and vacuum components). 

In the regime where the vacuum 
is allowed to cluster we have 
$\rho_{\Lambda s}(R)=\Lambda_{s}(R)/8\pi G$. Therefore, in such a 
situation it could
be possible, on non-linear scales, to have an interaction
between dark matter and dark energy with a different $\nu$
than the background value.
We also assume that the general functional form
that describes the behavior of the vacuum energy density
inside the spherical perturbation obeys a similar equation
as that of eq.(\ref{CRG}):
\begin{equation}
\label{CRGclust}
\Lambda_{s}(R)=\Lambda_0+
8\pi G \;\frac{\nu_{s}\,\rho_{ms, t}}{1-\nu_{s}}\,\left[
\left(\frac{R}{R_{t}}\right)^{-3(1-\nu_{s})}-1\right] \;.
\end{equation}
where $\nu_{s}$ is not necessarily equal to the background $\nu$.

Using the basic differential equations 
[see eq.(\ref{friedr}) and (\ref{rayc})]
and performing the 
following transformations $x=a/a_{t}$ and $y=R/R_{t}$  
we arrive at 
\begin{equation}
\label{xdot1}
{\dot x}^{2}=
H_{t}^{2}\Omega_{m,t} \left[x^{-1+3\nu}+rx^{2}I(x)\right]
\end{equation}
and
\begin{equation}\label{yddot}
\frac{{\ddot y}}{y}=-\frac{H_{t}^{2}\Omega_{ m,t}}{2} \left[
\frac{\zeta}{y^{3(1-\nu)}}-2\frac{\rho_{\Lambda s}}{\rho_{m,
t}}\right]\,.
\end{equation}
where $H^{2}_{t}\Omega_{ m,t}=\frac{8 \pi G}{3}\rho_{m, t}$. Notice that
$\Omega_{m,t}\equiv \Omega_{m}(a_{t})$ is the matter density 
parameter at the turnaround epoch (see eq.\ref{effeom}).

It is important to point that in order to obtain the above
set of equations we have used the following relations:
\begin{equation}
\rho_{ms}=\rho_{ms, t} \left(\frac{R}{R_{t}}\right)^{-3(1-\nu)}=
\frac{\zeta \rho_{m, t}}{y^{3(1-\nu)}} \;,
\end{equation}
\be
\label{Ixx}
I(x)=\frac{\rho_{\Lambda}}{\rho_{\Lambda, t}}=
\frac{1+\nu \tilde{r}_{0} a^{-3(1-\nu)}_{t} x^{-3(1-\nu)} }
{1+\nu \tilde{r}_{0} a^{-3(1-\nu)}_{t} } \;,
\ee
and 
\begin{equation}
\label{Irr}
r=\frac{\rho_{\Lambda, t}}{\rho_{m, t}}
=\frac{\Omega_{\Lambda}}{\Omega_{m}}a^{3(1-\nu)}_{t}+
\frac{\nu}{1-\nu}\left[1-a^{3(1-\nu)}_{t}\right]
\end{equation}
where $\tilde{r}_0=\tilde{\Omega}_{\Lambda}/\tilde{\Omega}_{m}$, 
$\rho_{m,t}$, $\rho_{\Lambda, t}$ are the matter and 
the vacuum energy density at the turnaround 
epoch which satisfies $\Omega_{\Lambda,t}=1- \Omega_{m,t}$
(for definition see eq.\ref{effeom}). Of course we need to say 
that in order  
to derive eqs.(\ref{Ixx}) and (\ref{Irr}) we have used 
eqs.(\ref{otran1}) and (\ref{CRG}).
The matter density in the spherical region at the turn around time 
is given with respect to the background matter density
at the same epoch $\rho_{m, t}$ as 
$\rho_{ms, t}=\zeta \rho_{m, t}$. 
The parameter $\zeta$
is the density contrast at the turnaround point. 
It is interesting to mention that in the case 
of the Einstein-de Sitter model
($\tilde{\Omega}_{m}=\Omega_{m}=1$ and $\nu=0$) the solution of
the system formed by eq.(\ref{xdot1}) and eq.(\ref{yddot}) reduces to
the well known value of the density contrast at the turnaround
point: $\zeta=\left(\frac{3 \pi}{4}\right)^{2}$, as it should.

For bound perturbations which do not expand forever, the time needed 
to re-collapse 
is twice the turn-around time, $t_{c}\simeq 2t_{t}$. Therefore, 
from eq.(\ref{frie456t}), we obtain the relation 
between the $\alpha_{c}$ and $\alpha_{t}$ 
${\rm sinh^{-1}}\left[\sqrt{ \tilde{r}_{0} \;a_{c}^{3(1-\nu)}}
\right]\simeq 2\;{\rm sinh^{-1}}\left[\sqrt{\tilde{r}_{0} \;
a_{t}^{3(1-\nu)}} \right]$ which is well approximated by
$z_{t}\simeq 1.523z_{c}+0.8$.  
Therefore, considering that 
clusters have virialized at the present epoch, $z_{c}\simeq 0$, the 
turnaround redshift is $z_{t} \simeq 0.8$. 
If we assume that galaxy clusters have formed close to the 
epoch of $z_{c}\sim 1.5$ then we obtain $z_{t} \sim 3$. 
To this end we verify that the ratio between the 
scale factors converges to the Einstein de Sitter value
$(a_{c}/a_{t})_{\Lambda_{RG}} \simeq (1+z_{t})/(1+z_{c})= 2^{2/3}$ 
at high redshifts owing to the 
fact that the matter component dominates
the Hubble expansion. 

In order to provide the virial theorem of the $\Lambda_{RG}$ model 
we need to generalize the Layzer-Irvine equation,
which describes the flow to virialization \cite{Peeb93}. 
If we take into account 
that the matter is exchanging energy with
the vacuum then the modified virial theorem becomes\cite{Basvir}
\be \label{lazer4}
(2-3\nu)T+(1-6\nu)(U_{G}-2U_{\Lambda})=0 
\ee 
where 
$U_{G}=3GM^{2}/5R$. 
The vacuum potential energy $U_{\Lambda}$ is written as
\begin{equation} \label{VV} 
U_{\Lambda}=\left\{ \begin{array}{cc} -\frac{\Lambda(a)MR^{2}}{10}
\;\;
       &\mbox{Homogeneous}\\
  -\frac{\Lambda_{0}MR^{2}}{10}+\frac{4\pi G\nu_{s}M \rho_{ms, t}R^{2}}{5(1-\nu_{s})}- 
\frac{4\pi G\nu_{s} M \rho_{ms, t}R^{-1+3\nu_{s}}}{(1-\nu_{s})(2+3\nu_{s})R^{-3(1-\nu)}_{t}} \;\;
       & \mbox{Clustered}
       \end{array}
        \right.
\end{equation}
where $M$ is the mass inside the spherical overdensity.
Utilizing the observational constraint $\nu=0.0048$, the 
deviation from the usual virial condition is $\sim 2-3\%$.
In this context, combining the virial theorem and the energy conservation 
($T_{c}+U_{G,c}+U_{\Lambda,c}= U_{G,t}+U_{\Lambda,t}$) 
at the collapse and at the turn around epochs we 
reach to the following condition: 
\be 
\label{EE}
q_{1}(\nu)U_{G,c}+q_{2}(\nu)U_{\Lambda,c}=U_{G,t}+U_{\Lambda,t},\;\;\;
q_{1}(\nu)=\frac{1+3\nu}{2-3\nu}, \;\;\; q_{2}(\nu)=\frac{4-15\nu}{2-3\nu} \ . 
\ee 
Clearly, for $\nu=0$ the above equations boil down to 
those of the $\Lambda$ model.

\subsubsection{Homogeneous vacuum}
In the case of homogeneous vacuum (hereafter $\Lambda_{RGH})$ model 
we have $\rho_{\Lambda s}(a)=\rho_{\Lambda}(a)=\Lambda(a)/8\pi G$. 
Within this framework, inserting eq.(\ref{Ixx}) into
eq.(\ref{yddot}), we obtain
\begin{equation}
\label{yddot1} {\ddot y}=-\frac{H_{t}^{2}\Omega_{ m,t}}{2}
\left[ \frac{\zeta}{y^{2-3\nu}}-2r y I(x)\right].
\end{equation}
The solution for $\zeta$ is provided only numerically 
by integrating the main system of differential equations,
(eqs.\ref{xdot1} and \ref{yddot1}), using the boundary
conditions: $({\rm  d} y/{\rm d} x)=0$ and $y=1$ at $x=1$.
However, we find that 
a reasonably accurate fitting formula for $\zeta$ is given by
\be \label{zetacl1}
\zeta \simeq \left(\frac{3\pi}{4}\right)^{2}
\Omega_{m,t}^{-\omega_{1}+\omega_{2}\Omega_{m,t}-\omega_{3}w(a_{t})},\;\;\;\;\;\;
(\omega_{1},\omega_{2},\omega_{3})=(0.79,0.26,0.06)
\ee
where
$w(a)=-1-\frac{\nu a^{3\nu}} {a^{3\nu}+\tilde{r}_{0}}$.
Using eqs.(\ref{VV}),(\ref{EE}) and $U_{G}=3GM^{2}/5R$ 
we can obtain a cubic equation that relates the ratio between the
virial $R_{c}$ and the turn-around outer radius 
$R_{t}$ the so called collapse factor ($\lambda=R_{c}/R_{t}$),
$q_{2}(\nu)n_{c}
\lambda^{3}-(2+n_{t})\lambda+2q_{1}(\nu)=0$,
where
\begin{equation}
n_{c, t}=\frac{\Lambda(a_{c, t})}{4\pi G \rho_{m, t} \zeta}=
n_{0}+
\frac{2\nu a^{3(1-\nu)}_{t}}{\zeta (1-\nu)}\;
\left[a^{-3(1-\nu)}_{c, t}-1\right]
\end{equation}
with
$n_{0}=\frac{2 \Omega_{\Lambda}
a^{3(1-\nu)}_{t} } {\Omega_{m} \zeta}$.
The viable solution ($0<\lambda <1)$ of the above cubic equation is
\begin{equation}
\lambda=-
\frac{2d^{1/3}}{3} {\rm cos} \left(\frac{\theta-2\pi}{3} \right),
\;\;\;\;\;\; d=\sqrt{x_{1}^{2}+x_{2}^{2}}, \;\;\;\;\; \theta={\rm cos^{-1}} (x_{1}/d) 
\end{equation}
where $x_{1}=-27q_{1}$, $x_{2}=-\frac{3\sqrt{3 \cal{D}}}{2}$ and
${\cal D}=4\;\frac{(2+n_{t})^{3}-27q^{2}_{1}q_{2}n_{c}}
{q^{3}_{2}n^{3}_{c}}$.  
Of course in the case of $\nu=0$ the above expressions
get the usual form for $\Lambda$ cosmology\cite{Lahav91,Bas03}
while for an Einstein-de Sitter model ($\Omega_{m}=1$)
we have $\lambda=1/2$.

We find that the collapse factor
lies in the range $\lambda \simeq 0.48-0.50$ in agreement with
previous studies\cite{Bas03,Maor05,Wang06,Horel05,Perc05,Basi07,Pace10}. 
In the inner panel of figure 2 we
plot the relative 
deviation (solid line) 
of the collapse factors
$\lambda_{\Lambda_{RGH}}(z_{c})$ for the current vacuum model 
with respect to the $\Lambda$ solution $\lambda_{\Lambda}(z_{c})$.
Obviously, the deviation from the $\Lambda$CDM case is small $\sim -2\%$.  
In the outer panel of figure 2 we present the evolution of the
density contrast at virialization (solid curve)
\begin{equation}\label{deltavir}
\Delta_{vir}=\frac{\rho_{ms, c}}{\rho_{m, c}}=
\frac{\zeta}{\lambda^{3}} \left(\frac{a_{c}}{a_{t}}\right)^{3} \;.
\end{equation}
We verify, that the density contrast
decreases with
the virialization redshift $z\equiv z_{c}$ and at very large redshifts 
it tends to the Einstein-de Sitter value
($\Delta_{vir}\sim 18\pi^{2}$), since 
the matter component dominates the cosmic fluid.
Following the notations of
Ref.~[\refcite{Wein03}], we obtain 
an accurate fitting formula to $\Delta_{vir}$
(within a physical range of cosmological parameters and for $z<2$)
\be
\label{aproxf}
\Delta_{vir}(a)\simeq 18\pi^{2}\left[1+\epsilon \;\Theta^{b}(a)\right], \;\;\;\;
\Theta(a)=\Omega^{-1}_{m}(a)-1,
\ee
where $\epsilon =0.40-23.2\nu+500\nu^{2}$ and $b=0.94+39.6\nu$.
From the right panel of figure 2 we observe  
that the virial density $\Delta_{vir}$ of the
$\Lambda_{RGH}$ model is somewhat lower with respect 
to that of $\Lambda$CDM model, namely the relative difference can reach 
up to $\sim -4\%$.  
As an example, assuming that 
clusters have formed prior to the epoch of 
$z_{c}\simeq 1.5$ ($z_{t}\sim 3$) 
we find $(\zeta,\Delta_{vir})_{\Lambda_{RGH}}\simeq (5.66,183.9)$
and $(\zeta,\Delta_{vir})_{\Lambda}\simeq (5.65,190)$.
To conclude this discussion we would like to stress 
that in the homogeneous case the 
dark energy (in our case vacuum) 
component flows progressively out of the overdensity \cite{Maor05,david}
and hence energy conservation cannot be applied.  
Such violation is however only important 
for large values of $|\nu|\simeq {\cal O}(10^{-2})$ and 
at very late times, when the vacuum energy dominates the 
cosmic fluid. 

\subsubsection{Clustered vacuum}
In this section we assume that the vacuum clusters along  
with the dark matter. As we have already mentioned the 
vacuum inside the spherical patch is given 
by eq.(\ref{CRGclust}). 
In the current study we restrict our analysis to 
$\nu_{s}=0.002$ (hereafter $\Lambda_{RGC1}$: dashed line) and
$\nu_{s}=-0.002$ (hereafter $\Lambda_{RGC2}$: dotted line). 
Substituting $y=R/R_{t}$ into eq.(\ref{CRGclust}) we have 
\begin{equation}
\label{CRGclust1}
\Lambda_{s}(y)=\Lambda_0+
8\pi G \;\frac{\nu_{s}\,\zeta \rho_{m, t}}{1-\nu_{s}}\,
\left[y^{-3(1-\nu_{s})}-1\right] \;,
\end{equation}
where the vacuum energy density is 
$\rho_{\Lambda s}(y)=\Lambda_{s}(y)/8\pi G$. 
The merit of the latter assumption is that it allows 
an analytical solution to the system of eqs.(\ref{xdot1}) and (\ref{yddot}).
Indeed inserting 
the above form of $\rho_{\Lambda s}(y)$ into eq (\ref{yddot}) we obtain  
\begin{equation}
\label{yddotcl} {\ddot y}=-\frac{H_{t}^{2}\Omega_{m,t}}{2}
\left[ \frac{(1-3\nu_{s})\zeta}{(1-\nu_{s})y^{2-3\nu_{s}}}-
2\left(r-\frac{\nu_{s} \zeta}{1-\nu_{s}}\right)y \right]\;.
\end{equation}
and upon integration we find
\begin{equation}
\label{yddotcl2} 
{\dot y}^{2}=H_{\rm t}^{2}\Omega_{m, t}\left[P(y,\zeta)+C \right],
\end{equation}
where $C$ is the integration constant and
$P(y,\zeta)=\frac{\zeta}{(1-\nu_{s})y^{1-3\nu_{s}}}+\left(r-\frac{\nu_{s} \zeta}{1-\nu_{s}}\right)y^{2}$.
Notice that the boundary conditions, $(dy/dx)=0$
and $y=1$ at $x=1$, imply that $C=-P(1,\zeta)$. Combining 
eq.(\ref{yddotcl2})  
with the background equation (\ref{xdot1}) the solution of the system is
\begin{equation}
\label{zzecl}
\int_{0}^{1} \frac{dy}{\sqrt{P(y,\zeta)-P(1,\zeta)}}=
\int_{0}^{1} \frac{dx}{\sqrt{x^{-1+3\nu}+rx^{2}I(x)}} \;.
\end{equation}
Furthermore, if the vacuum energy participates in the virialization then 
the potential energy of the vacuum is given by the second branch 
of eq.(\ref{VV}). In this case, 
using simultaneously eqs.(\ref{VV}),(\ref{EE}) and $U_{G}=3GM^{2}/5R$ 
the collapse factor obeys 
\begin{equation}
\label{collapcl}
q_{2}(\nu_{s})[n_{0}-f(\nu_{s})]\lambda^{3}-A(n_{0},\nu_{s})\lambda
+g(\nu_{s})\lambda^{3\nu_{s}}+2q_{1}(\nu_{s})=0\,,
\end{equation}
where 
$$f(\nu_{s})=\frac{2\nu_{s}}{1-\nu_{s}},\;\;
g(\nu_{s})=\frac{10\nu_{s}q_{2}(\nu_{s})}{(1-\nu_{s})(2-3\nu_{s})},\;\;
A(n_{0},\nu_{s})=2+n_{0}-f(\nu_{s})+\frac{g(\nu_{s})}{q_{2}(\nu_{s})}.$$
Finally, solving eqs.(\ref{zzecl}) 
and (\ref{collapcl}), we can estimate
the density contrast at virialization 
from eq.(\ref{deltavir}).
Similarly, as in section 6.2.1, the fitting formulas
for $\zeta$ as well as for $\Delta_{vir}$ are given by 
eq.(\ref{zetacl1}) and eq.(\ref{aproxf}).
The corresponding parameters of the approximated 
formulas are 

\begin{equation}
(\omega_{1},\omega_{2},\omega_{3})=\left\{ \begin{array}{cc}
       (0.62,-0.08,0.06)\;\;\;\;
       0\le \nu_{s}\le 0.003 \\
(0.82,0.22,0.06)\;\;\;\;\; -0.003\le \nu_{s}< 0
       \end{array}
        \right.
\end{equation}

\begin{equation}
(b,\epsilon)=\left\{ \begin{array}{cc}
       (0.94+90\nu_{s},0.40-46\nu_{s}+500\nu_{s}^{2}) \;\;\;
       -0.002< \nu_{s}\le 0.003 \\
       (0.94+95\nu_{s},0.31-129\nu_{s}+500\nu_{s}^{2})\;\;\; -0.003\le \nu_{s}\le -0.002

       \end{array}
        \right.
\end{equation}
In this framework 
the collapse factor obeys $0.46 \le \lambda \le 0.52$. 
Also from figure 2
(see inner and outer panels) we see that 
the largest positive deviation of the
collapse factor occurs for the $\Lambda_{RGC2}$ model (dotted line).
This implies that $\Lambda_{RGC2}$ model forms more bound
systems than the concordance $\Lambda$CDM model (solid points) and thus 
the corresponding cosmic structures should
be located in larger density environments. 
Indeed, at the cluster formation epoch 
$z_{c}\simeq 1.5$ we obtain 
$(\zeta,\Delta_{vir})_{\Lambda_{RGC2}}\simeq (5.66,202)$. The opposite
situation holds for the $\Lambda_{RGC1}$ model. 
From figure 2 it becomes obvious 
that the size and $\Delta_{vir}$ of the
cosmic structures which are produced in the $\Lambda_{RGC1}$ model (dashed line)
are remarkably close to that predicted by the $\Lambda_{RGH}$ vacuum 
cosmology, and therefore the impact of the 
vacuum energy on the spherical collapse is very small in the
clustered case as long as $\nu_{s}$ is positive. 
In other words we find that both $\Lambda_{RGH}$ and 
$\Lambda_{RGC1}$ models are equivalent at the background and perturbation
levels.
Finally, at the epoch of $z_{c}\simeq 1.5$ we find 
$(\zeta,\Delta_{vir})_{\Lambda_{RGC1}}\simeq (5.70,184.2)$. 
For comparison we provide the density pair of the 
$\Lambda_{RGH}$ model, namely 
$(\zeta,\Delta_{vir})_{\Lambda_{RGH}}\simeq (5.66,183.9)$.

\section{Conclusions}
In this review article 
we have studied the overall dynamics of the FLRW flat cosmological models 
in which the vacuum energy varies with the Hubble parameter, namely 
$\Lambda(H)=\Lambda_{0}+3\nu(H^{2}-H^{2}_{0})$. First we have performed a joint
likelihood analysis in order to put constraints on the main
cosmological parameters by using the current observational data
(SNIa, BAOs and CMB shift parameter together with the growth rate
of galaxy clustering). 
We have shown that the $\Lambda(H)$ model fits slightly better 
the observational data than that of the traditional $\Lambda$ cosmology. 
In particular, we have found that the $\Lambda$CDM model 
can not simultaneously accommodate the 
{\em Planck} priors and the growth data implying that 
this kind of data favor the $\Lambda(H)$ vacuum scenario.
Subsequently we have investigated the nonlinear regime
and considered the predicted redshift 
distribution of cluster-size collapsed structures as a
powerful method to distinguish the $\Lambda(H)$ and 
$\Lambda$CDM cosmological scenarios.
Finally, we have generalized the properties 
(virial theorem, collapse factor, virial and turnaround densities) 
of the spherical collapse model 
in the case when the vacuum energy is a running function of
the Hubble rate, $\Lambda=\Lambda(H)$. 
Overall, we have found 
that the virial density contrast is affected by the
considered status of the vacuum energy model (homogeneous or
clustered). 

{\bf Acknowledgments:}
I would like 
to thank J. Sol\`a, M. Plionis, J. A. S. Lima 
D. Polarski, N. E. Mavromatos, 
L. Perivolaropoulos and A. Gomez-Valent
for the recent collaboration in some of the work presented in this review 
article.

\end{document}